\documentclass{article}

\usepackage[final]{neurips_2019_ml4ps}

\usepackage[utf8]{inputenc} 
\usepackage[T1]{fontenc}    
\usepackage{hyperref}       
\usepackage{url}            
\usepackage{booktabs}       
\usepackage{amsfonts}       
\usepackage{nicefrac}       
\usepackage{microtype}      

\usepackage{graphicx}
\usepackage{subfig}

\title{From Dark Matter to Galaxies with Convolutional Neural Networks}

\author{
  Jacky H. T. Yip\thanks{These authors contributed equally.} \\
  Department of Physics, The Chinese University of Hong Kong \\
  \texttt{1155092406@link.cuhk.edu.hk} \\
  \And
  Xinyue Zhang\footnotemark[1], Yanfang Wang\footnotemark[1], Wei Zhang\footnotemark[1], Yueqiu Sun\footnotemark[1] \\
  Center for Data Science, New York University \\
  \And
  Gabriella Contardo, Francisco Villaescusa-Navarro, Siyu He, Shy Genel, Shirley Ho \\
  Center for Computational Astrophysics, Flatiron Institute \\ 
}

\begin{document}

\maketitle

\begin{abstract}
Cosmological simulations play an important role in the interpretation of astronomical data, in particular in comparing observed data to our theoretical expectations. However, to compare data with these simulations, the simulations in principle need to include gravity, magneto-hydrodyanmics, radiative transfer, etc. These ideal large-volume simulations (\textit{gravo-magneto-hydrodynamical}) are incredibly computationally expensive which can cost tens of millions of CPU hours to run. In this paper, we propose a deep learning approach to map from the dark-matter-only simulation (computationally cheaper) to the galaxy distribution (from the much costlier cosmological simulation). The main challenge of this task is the high sparsity in the target galaxy distribution: space is mainly empty. We propose a cascade architecture composed of a classification filter followed by a regression procedure. We show that our result outperforms a state-of-the-art model used in the astronomical community, and provides a good trade-off between computational cost and prediction accuracy.
\end{abstract}

\section{Introduction}
Most cosmological surveys observe galaxies, which are however very difficult to model directly due to the complicated physics involved in their formation and evolution. For even a small fraction of the Universe, evolving tens of billions of resolution particles interacting under coupled effects of gravity, magneto-hydrodynamics and radiative processes over cosmic time is incredibly computationally costly; a state-of-the-art simulation today, such as IllustrisTNG [1], requires up to 40 million CPU hours to complete ($\sim$4500 years on a single CPU).

On the other hand, in the $\Lambda$CDM cosmological model, baryonic matter ("normal" matter, which galaxies are made out of) constitutes only about one-sixth of all the matter in the Universe. The rest is composed of (notably) dark matter, that interacts only through gravity and drives the growth and morphology of the large-scale structure of the Universe, such as galaxy clusters, filaments, and voids. Baryonic matter collapses within dark matter halos, which results in the formation of stars and galaxies. In other words, dark matter halos are the environments in which galaxies form, evolve and merge. Hence, the behaviors and properties of galaxies, such as their spatial distribution, are expected to be closely connected to the properties of dark matter halos they live in.

In contrast to full hydrodynamical simulations, dark-matter-only N-body simulations are computationally much cheaper as gravity is the only interacting force. Therefore, it would be extremely interesting to find the mapping between the dark matter and galaxy fields in N-body and full hydrodynamical simulations, respectively. State-of-the-art  Halo Occupation Distribution (hereafter HOD) [3] model has been developed to achieve this, but it is unclear whether it is comprehensive enough as the model usually relies on certain assumptions such as the halo mass being the main quantity controlling galaxy properties, and not, for instance, its local structure.

We propose a first deep learning approach for the above purpose. We explore the use of a cascade of convolutional neural networks (CNNs) to map from the dark matter distribution, obtained from a gravity-only N-body simulation, to the galaxy distribution, obtained from a full gravo-magneto-hydrodynamical simulation. We show that our model predicts a significantly more reliable galaxy distribution compared to the HOD algorithm on a variety of criteria, including statistics commonly used in cosmology.

\setcounter{footnote}{0}

\section{Method}
\subsection{Data}
From the IllustrisTNG project [2], we use the TNG300 full hydrodynamical simulation (TNG300-1) since it simulates the largest range of spatial scales for this type of simulations. We also employ the associated N-body simulation (TNG300-1-Dark). We use the level-1 simulations (highest spatial and mass resolution) at present day Universe (redshift $z=0$). Note that our approach could be applied in similar fashion on any simulation pairs of dark matter and their respective hydrodynamical simulation.

The input data is the mass density field of the dark matter distribution, which allows us to be independent of the specific configurations of the simulation such as the mass resolution. The target data is the galaxy number density field. We define galaxies as the subhalos with a positive stellar mass (total mass of stars particles)\footnote{The \texttt{SubhaloFlag} field from the catalog is also used to filter out non-cosmological subhalos.}. In the following experiments, we do not apply a threshold on the minimum stellar mass to our target set of galaxies, i.e. we will predict all galaxies with non-zero stellar mass.

Both the dark matter particles and the galaxies from the TNG300 simulations are enclosed in a 3D simulation volume of (205 Mpc/h)$^3$. We grid this space into 1024$^3$ voxels, and generate sub-cubes (taken as inputs and targets) of 32$^3$ voxels each. The resulting dataset is composed of 32$^3$ pairs of samples. We split this dataset into training (63.1\%), validation (19.1\%) and test (17.8\%) sets. The test set forms a coherent volume of (115.3 Mpc/h)$^3$ for computing the relevant cosmological statistics. 

\subsection{Cascade architecture}
We physically expect that the formation of a galaxy should primarily depend on the local properties of the halo and the environment where it lives, and that it is independent of its absolute position. This motivates our choice of convolutional neural network (CNN) based architecture [6]. However, the percentage of non-empty voxels is 75.93\% for the input field and 0.15\% for the target galaxy field. The high sparsity in the latter is the major difficulty of the task, because the model would still achieve a high accuracy (99.85\%) even if it fails to predict all the galaxies. We thus propose a two-phase cascade architecture to overcome this problem.
  
\paragraph{Classification phase.} 
The first phase is an \textit{Inception Network} [4] as a classifier to predict the probability of having at least one galaxy in each of the target voxels. We use the weighted cross-entropy loss for this binary classification problem:
\begin{eqnarray}\mathbb{L}_{\mathrm{CrossEnt}}(\mathbf{x},\mathbf{y})=-mean\{[w_1\cdot\mathbf{y}\cdot\mathrm{log}(\mathbf{x})+(1-\mathbf{y})\cdot\mathrm{log}(1-\mathbf{x})]/(w_1+1)\}~\end{eqnarray}
where $\mathbf{x}$ contains the predicted probabilities in the voxels for the presence of galaxies, $\mathbf{y}$ contains the target values (1 for presence; 0 for absence) and ${w_1}$ is the weight applied to emphasize the penalty on bad predictions in the voxels in which galaxies are present. This weight effectively corrects the imbalance in the target classes due to the high sparsity. 
  
\paragraph{Regression phase.}
The second phase is a \textit{Recurrent Residual U-Net} [5]. Similarly, it takes the dark matter density field as input, and regresses the number of galaxies $\mathbf{n_g}$ in each voxel. The output from the previous phase is used as a mask to compute the loss and backpropagate when considering only the voxels that were predicted in the previous phase to have at least one galaxy. We optimize this phase using a weighted mean square error (MSE) loss as follows:
\begin{eqnarray}\mathbb{L}_{\mathrm{MSE}}(\mathbf{n_g},\mathbf{n_t}) = mean\{W(\mathbf{n_t})(\mathbf{n_g} - \mathbf{n_t})^2\}\end{eqnarray}
where $\mathbf{n_t}$ contains the numbers of galaxies in the target, and the weight function $W$ is defined by $W(n_t,_i)=w_2$ for $n_t,_i>0$ and $W(n_t,_i)=1$ otherwise. Similarly, this weight function provides a means (by varying $w_2$) to emphasize the loss from the voxels in which galaxies are present in the target.

The output of this cascade model is continuous (not categorical integer prediction), which can be interpreted as a probabilistic number of galaxies in each voxel. As we focus here on statistics which do not require exact numbers of galaxies (e.g. the overdensity field used for computing the power spectrum and bispectrum in the next Section), we keep the output as is. Otherwise, one may round the output to obtain a generic number density field. 

\paragraph{Hyperparameter search and training.}
While the MSE (Eqn. (2)) is effective as the loss, it is not an ideal indicator of the model's performance: if the model predicts a galaxy's presence in a voxel next to where it truly is, this MSE would yield a significant error while the slightly misplaced galaxy prediction does not practically make a big difference in most statistics of interest here. Therefore, for determining which model performs best, we use the MSE in the total number of galaxies in each 32$^3$ sub-cube (MSE$_{sub}$), which is the baseline statistic that should be matched well by the prediction.

The classifier and the regression model are trained separately. We prioritize training the classifier for a high recall in order to minimize the number of non-empty voxels incorrectly removed (false negative) as they would become unrecoverable in the regression phase. For instance, with $w_1=500$, we obtain a result with 99.07\% recall and 8.28\% precision, i.e. only $\sim$1\% of non-empty voxels are incorrectly classified while the sparsity is effectively reduced by $\sim$55 times. For the regression model, the weight $w_2$ is tuned to minimize the corresponding MSE$_{sub}$ on the validation set. The optimal weight is found to be $w_2=1.05$.

The model in each phase is trained for 50 epochs with the Adam optimizer and a batch size of 16. The learning rate starts from 0.001 and is halved every 4 epochs. We augment the training data by using random rotations of the cubes. All hyper-parameters are selected based on the performance on the validation set. The total training time with 1 GPU is $\sim$10 hours.

\section{Results}
All results shown here are computed on the test set. We compare our model's result with the state-of-the-art Halo Occupation Distribution model. The HOD model identifies mass-related-only parameters to determine the number of galaxies that a dark matter halo holds, then the galaxies are placed randomly within a critical radius inside the halo.

\paragraph{Visualization.} Figure 1 shows snapshots of a slice from the input, target and results of the cascade model and HOD. Both the cascade model and HOD successfully predict approximate positions of the galaxies. However, from the zoomed-in images (second row), it is evident that our cascade model (third column) learns small scale details on the distribution of galaxies, while HOD only predicts a spherical distribution of galaxies within the halo.

\paragraph{Power spectra and bispectra.} Summary statistics such as the power spectrum and the bispectrum are most commonly used in cosmology to extract information from fields of fluctuation, such as the overdensity field of the galaxy distribution. The power spectrum $P(k)$ is the Fourier equivalent of the two-point correlation function which measures how the actual galaxy distribution deviates from a simple Gaussian random field. On the other hand, the bispectrum $B(k)$ is the Fourier equivalent of the three-point correlation function, which is a higher-order statistic commonly used for characterizing the galaxy distribution on smaller scales as a non-Gaussian field due to non-linear interactions. Figure 2 shows the power spectra (left) and bispectra (right) of the target and the results of our model and HOD.

For the power spectrum, our cascade model obtains a good fit on a large range of scales even though it is trained on relatively small sub-cubes. Over the full range of scales, its mean relative residual\footnote{The mean relative residual is defined by $mean\{|\mathbf{y_m}/\mathbf{y_t}-1|\}$, where $\mathbf{y_m}$ and $\mathbf{y_t}$ are spectrum values of the model's prediction and the target respectively.} is 26.9\% which is much smaller than HOD's 246.6\%. Our model achieves a comparable performance as HOD for $k<0.3$ h/Mpc (larger scales), and outperforms for larger $k$ (smaller scales). This shows that our model can better capture the non-linearities in the smaller scales of the field. This is further demonstrated in the bispectra on small scales ($k_1=1.2$ h/Mpc and $k_2=1.3$ h/Mpc), in which our model outperforms HOD by a significant margin (0.79\% vs 435\% in mean relative residual).

\begin{figure}[ht!]
\vspace{-0.1375cm}
  \centering
  \includegraphics[clip,trim=60 40 0 40 ,scale=0.3]{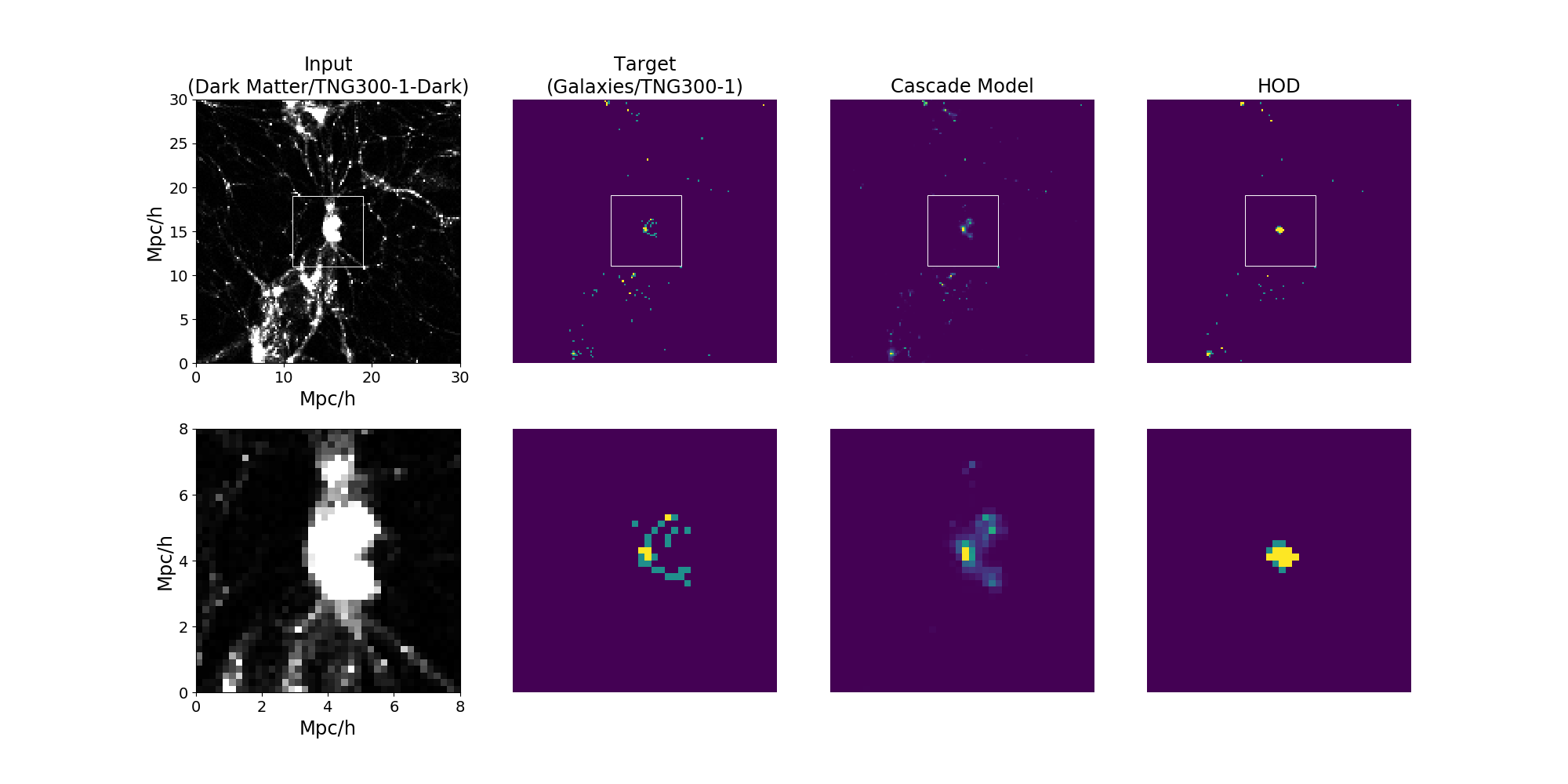}
  \caption{Snapshots of a slice from simulations and results: First column shows the dark matter input; Second shows the target galaxies; Third shows the prediction from our cascade model; Forth is from our benchmark model, the commonly deployed method in cosmology. Second row is a zoomed-in on the white squares in the first row.  Brighter colors represent more dark matter particles/galaxies\protect\footnotemark.}
\end{figure}
\footnotetext{For illustration purposes, colors in the galaxies snapshots range from blue-violet (darkest) to blue-green and to yellow-orange (brightest). Yellow-orange represents voxels with 2 or more galaxies, blue-green represents voxels with 1 galaxy and blue-violet represents empty voxels in the backgrounds. For our cascade model's prediction, there are more intermediate colors as it predicts continuous numbers of galaxies.}
\begin{figure}[ht!]
  \vspace{-0.725cm}
  \centering
  \subfloat{{\includegraphics[clip,trim=12 190 60 55, scale=0.32]{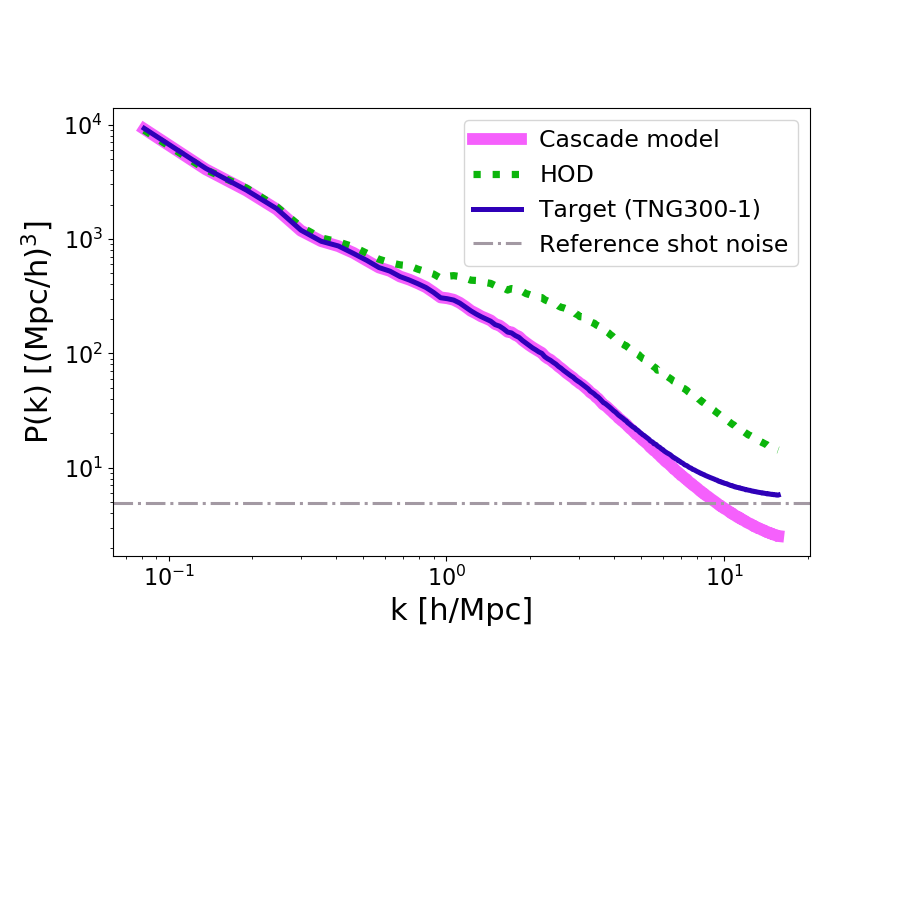}}}
  \qquad
  \subfloat{{\includegraphics[clip,trim=22 190 60 55, scale=0.32]{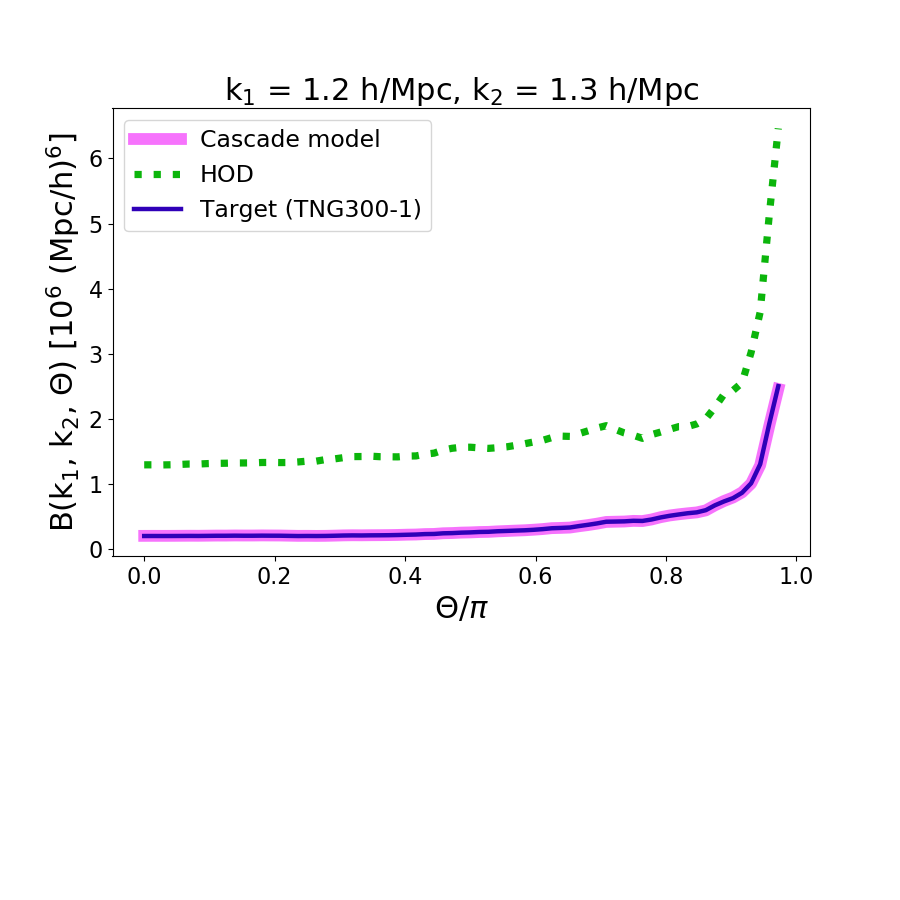}}}
  \caption{Power spectra $P(k)$ (left) and bispectra $B(k_1, k_2, \theta)$ (right)}
\end{figure}

\section{Conclusion and future work}
We show in this paper that our cascade model of convolutional neural networks can efficiently predict the number density field of galaxies given only the dark matter density field, and can outperform a benchmark method used in cosmology in a large range of scales.
  
We plan for several extensions of this work, notably focusing on predicting additional properties of the galaxies (e.g. stellar mass, star formation rate, etc). We are also looking into the ability of the model to transfer between simulations of different volumes and resolutions. Not only this work paves the way for obtaining simulation results more efficiently, but also could help explore scientific questions. For example, we are interested in whether our model can learn about halo assembly bias [7], that HOD neglects. Our approach could be used to analyse any possible effects of the environment (e.g. structure of the dark matter halos) regarding galaxy formation.
  
\subsubsection*{Acknowledgments}
We thank David Spergel, Siamak Ravanbakhsh and Barnabas Poczos for insightful discussions. This project is supported by the Center for Computational Astrophysics of the Flatiron Institute in New York City. The Flatiron Institute is supported by the Simons Foundation.
  
\section*{References}
\small
[1] Annalisa Pillepich, Volker Springel, Dylan Nelson, Shy Genel, Jill Naiman, Ruediger Pakmor, Lars Hernquist, Paul Torrey, Mark Vogelsberger, Rainer Weinberger, Federico Marinacci. Simulating Galaxy Formation with the IllustrisTNG Model.  \textit{Mon. Notices Royal Astron. Soc} 473 3 4077–4106, 2018

[2] Dylan Nelson, Volker Springel, Annalisa Pillepich, Vicente Rodriguez-Gomez, Paul Torrey, Shy Genel, Mark Vogelsberger, Ruediger Pakmor, Federico Marinacci, Rainer Weinberger, Luke Kelley, Mark Lovell, Benedikt Diemer, Lars Hernquist. The IllustrisTNG Simulations: Public Data Release. \textit{ArXiv e-prints}, Apr 2019

[3] Andreas A. Berlind, David H. Weinberg. The Halo Occupation Distribution: Towards an Empirical Determination of the Relation Between Galaxies and Mass. \textit{Astrophys. J.} 575 587-616, 2002

[4] Christian Szegedy, Wei Liu, Yangqing Jia, Pierre Sermanet, Scott Reed, Dragomir Anguelov, Dumitru Erhan, Vincent Vanhoucke, Andrew Rabinovich. Going Deeper with Convolutions. \textit{ArXiv e-prints}, Sep 2014

[5] Md Zahangir Alom, Mahmudul Hasan, Chris Yakopcic, Tarek M. Taha, Vijayan K. Asari. Recurrent Residual Convolutional Neural Network based on U-Net (R2U-Net) for Medical Image Segmentation. \textit{ArXiv e-prints}, May 2018

[6] Stéphane Mallat. Understanding deep convolutional networks. \textit{Philos. Trans. Royal Soc. A Math. Phys. Sci.} 374 2065, Apr 2016

[7] Mohammadjavad Vakili, ChangHoon Hahn. How are galaxies assigned to halos? Searching for assembly bias in the SDSS galaxy clustering. \textit{Astrophys. J.} 872 1, 2019

\end{document}